# Weyl Projective Curvature Symmetry in FRW k=0 Model

Ghulam Shabbir

Faculty of Engineering Sciences, GIK Institute of Engineering
Sciences and Technology, Topi Swabi, NWFP, Pakistan

Email: shabbir@giki.edu.pk

**Abstract**

A study of proper Weyl projective curvature collineations in FRW k=0 space-time is given using the rank of $6 \times 6$ Weyl projective curvature matrix and direct integration techniques. It is shown that a very special class of the above space-time admits proper Weyl projective curvature collineation.

## 1. INTRODUCTION

In spite of much interest in the study of symmetry in general relativity theorey, Weyl projective curvature symmetry seems to be ignored because of complications involved in (5). The Weyl projective curvature symmetry which preserve the Weyl projective curvature tensor carries significant interest in Einstein's theory of general relativity. It is therefore important to study these symmetries. In this paper a study of FRW k=0 model is given by using the rank of $6 \times 6$ Weyl projective curvature matrix and direct integration techniques.

Through out $M$ is representing the four dimensional, connected, hausdorff space-time manifold with Lorentz metric $g$ of signature (-, +, +, +). The curvature tensor associated with $g_{ab}$, through Levi-Civita connection, is denoted in component form by $R^a{}_{bcd}$, the Ricci tensor components are $R_{ab} = R^c{}_{acb}$, the Weyl tensor components are $C^a{}_{bcd}$, and the Weyl projective curvature tensor components are $W^a{}_{bcd}$. The usual covariant, partial and Lie derivatives are denoted by a semicolon, a comma and the symbol $L$, respectively. Round and square brackets denote the usual symmetrization and skew-symmetrization, respectively. Here, $M$ is assumed non-flat in the sense that the Weyl projective curvature tensor does not vanish over any non-empty open subset of $M$.

Any vector field $X$ on $M$ can be decomposed as

$$X_{a;b} = \frac{1}{2} h_{ab} + F_{ab} \tag{1}$$

where $h_{ab}(=h_{ba}) = L_X g_{ab}$ and $F_{ab}(=-F_{ba})$ are symmetric and skew symmetric tensor on $M$, respectively. Such a vector field $X$ is called projective if the local diffeomorphisms $\psi_t$ (for appropriate $t$) associated with $X$ map geodesics into geodesics. This is equivalent to the condition that $h_{ab}$ satisfies

$$h_{ab;c} = 2g_{ab}\eta_c + g_{ac}\eta_b + g_{bc}\eta_a \tag{2}$$

for some smooth closed 1-form on $M$ with local components $\eta_a$. Thus $\eta_a$ is locally gradient because the connection is metric and will, where appropriate, be written as $\eta_a = \eta_{,a}$ for some function $\eta$ on some open subset of $M$. If $X$ is a projective collineation and $\eta_{a;b} = 0$ then $X$ is called a special projective collineation on $M$. The statement that $h_{ab}$ is covariantly constant on $M$ is, from (2), equivalent to $\eta_a$ being zero on $M$ and is, in turn equivalent to $X$ being an affine vector field on $M$ (so that the local diffeomorphisms $\psi_t$ preserve not only geodesics but also their affine parameters). If $X$ is projective but not affine then it is called proper projective collineation [2]. Further if $X$ is affine and $h_{ab} = 2cg_{ab}, c \in R$ then $X$ is homothetic (otherwise proper affine). If $X$ is homothetic and $c \neq 0$ it is proper homothetic while if $c = 0$ it is Killing. A vector field $X$ on $M$ is called a curvature collineation (CC) if it satisfies [1]

$$L_X R^a{}_{bcd} = 0. \tag{3}$$

$X$ is said to be proper CC if it is not affine. At this point it is convenient to introduce the Weyl projective curvature tensor $W$ defined by

$$W^a{}_{bcd} = R^a{}_{bcd} - \frac{1}{3}\left(\delta^a_c R_{bd} - \delta^a_d R_{bc}\right) \tag{4}$$

A vector field $X$ on $M$ is called a Weyl projective curvature collineation (WPCC) if it satisfies [1]

$$L_X W^a{}_{bcd} = 0 \tag{5}$$

or equivalently

$$W^a{}_{bcd;e} X^e + W^a{}_{ecd} X^e{}_{;b} + W^a{}_{bed} X^e{}_{;c} + W^a{}_{bce} X^e{}_{;d} - W^e{}_{bcd} X^a{}_{;e} = 0.$$

One can expand the above equation (5) in a set of 22 coupled WPCC equations

$$W^a{}_{1a1} X^1{}_{,0} + W^b{}_{0b0} X^0{}_{,1} = 0, \quad (a,b) = (0,1),(2,2),(3,3) \tag{6}$$

$$W^a{}_{2a2} X^2{}_{,0} + W^b{}_{0b0} X^0{}_{,2} = 0, \quad (a,b) = (0,2),(1,1),(3,3) \tag{7}$$

$$W^a{}_{3a3} X^3{}_{,0} + W^b{}_{0b0} X^0{}_{,3} = 0, \quad (a,b) = (0,3),(1,1),(2,2) \tag{8}$$

$$W^a{}_{1a1,e} X^e + 2 W^a{}_{1a1} X^1{}_{,1} = 0, \quad a = 0,2,3 \text{ and } e = 0 \text{ or } 1 \tag{9}$$

$$W^a{}_{1a1} X^1{}_{,2} + W^b{}_{2b2} X^2{}_{,1} = 0, \quad (a,b) = (0,0),(2,1),(3,3) \tag{10}$$

$$W^a{}_{1a1} X^1{}_{,3} + W^b{}_{3b3} X^3{}_{,1} = 0, \quad (a,b) = (0,0),(3,1),(2,2) \tag{11}$$

$$W^a{}_{2a2} X^2{}_{,3} + W^b{}_{3b3} X^3{}_{,2} = 0, \quad (a,b) = (0,0),(1,1),(3,2) \tag{12}$$

$$W^a{}_{2a2,e} X^e + 2 W^a{}_{2a2} X^2{}_{,2} = 0, \quad a = 0,1,3 \text{ and } e = 0 \text{ or } 1 \tag{13}$$

$$W^a{}_{3a3,e} X^e + 2 W^a{}_{3a3} X^3{}_{,3} = 0, \quad a = 0,1,2 \text{ and } e = 0 \text{ or } 1 \tag{14}$$

$$W^a{}_{0a0,e} X^e + 2 W^a{}_{0a0} X^0{}_{,0} = 0, \quad a = 1,2,3 \text{ and } e = 0 \text{ or } 1 \tag{15}$$

$$(W^0{}_{a0a} - W^b{}_{aba}) X^0{}_{,b} = 0, \quad (a,b) = (1,3),(2,3),(1,2),(3,2),(2,1),(3,1) \tag{16}$$

$$(W^0{}_{a0a} - W^b{}_{aba}) X^b{}_{,0} = 0, \quad (a,b) = (1,3),(2,3),(1,2),(3,2),(2,1),(3,1) \tag{17}$$

$$(W^b{}_{aba} - W^c{}_{aca}) X^b{}_{,c} = 0, \quad (a,b,c) = (0,1,2),(3,1,2),(0,1,3),(2,1,3) \tag{18}$$

$$(W^b{}_{aba} - W^c{}_{aca}) X^c{}_{,b} = 0, \quad (a,b,c) = (0,1,2),(3,1,2),(0,3,2),(1,3,2) \tag{19}$$

$$(W^b{}_{aba} - W^c{}_{aca}) X^c{}_{,b} = 0, \quad (a,b,c) = (0,1,3),(2,1,3),(0,2,3),(1,2,3). \tag{20}$$

It is important to note that there is no summation over the repeated indices from equations (6) to (20). *Finally, X is said to be proper Weyl projective curvature collineation if it is not curvature collineation and projective collineation.*

## 2. Main Result

Consider the FRW k=0 space-time in the usual coordinate system $(t,x,y,z)$ (labeled by $(x^0, x^1, x^2, x^3)$, respectively) with line element

$$ds^2 = -dt^2 + R^2(t)(dx^2 + dy^2 + dz^2). \tag{21}$$

The above space-time admits six independent Killing vector fields which are

$$\frac{\partial}{\partial x}, \frac{\partial}{\partial y}, \frac{\partial}{\partial z}, x\frac{\partial}{\partial y} - y\frac{\partial}{\partial x}, z\frac{\partial}{\partial y} - y\frac{\partial}{\partial z}, x\frac{\partial}{\partial z} - z\frac{\partial}{\partial x}. \tag{22}$$

The Ricci tensor Segre type of the above space-time is $\{1,(111)\}$ or $\{(1,111)\}$. If the Segre type is $\{(1,111)\}$ then the above space-time is of constant curvature and the Weyl projective curvature tensor becomes zero [3]. The Petrvo type of the above space-time is O. The situations are well known when the above space-time (21) admits proper curvature collineation [4,5], proper projective vector field [1,4], proper affine vector field [6] and proper homothetic vector field [7,8]. These symmetries occur in special cases and some of them will be explained later. The non-zero independent components of the Weyl projective curvature tensor are

$$W^{01}{}_{01} = W^{02}{}_{02} = W^{03}{}_{03} = \frac{1}{3R^2}(4R\ddot{R} + 2\dot{R}^2) \equiv \alpha$$

$$W^{12}{}_{12} = W^{13}{}_{13} = W^{23}{}_{23} = \frac{1}{3R^2}(R\ddot{R} + 5\dot{R}^2) \equiv \beta,$$

The Weyl projective curvature tensor of $M$ can be described by components $W^{ab}{}_{cd}$ written as a $6\times 6$ symmetric matrix. They take the form [8]

$$W^{ab}{}_{cd} = diag(\alpha, \alpha, \alpha, \beta, \beta, \beta) \tag{23}$$

where $\alpha$ and $\beta$ are real functions of $t$ only and the six dimensional labelling is in the order $01, 02, 03, 12, 13, 23$ with $x^0 = t.$ Since we are interested in the rank of the non flat $6\times 6$ Weyl projective curvature matrix, so it follows from (23) that there exists the following possibilities:

(A)   Rank=6, $\alpha \neq 0$ and $\beta \neq 0$

(B)   Rank=3, $\alpha \neq 0$ and $\beta = 0$

(C)   Rank=3, $\alpha = 0$ and $\beta \neq 0$.

We consider each case in turn.

**Case A**

In this case we have $\alpha \neq 0$, $\beta \neq 0$ and the rank of the $6\times 6$ Weyl projective curvature matrix is six. Subsitituting the above information into the Weyl projective curvature equations (6) to (20) and after lenghty calculation one finds that there exists they following two subcases which are

(a)  $\dot{R}^2 - R\ddot{R} = 0$  (b)  $\dot{R}^2 - R\ddot{R} \neq 0$.

We will consider each subcase in turn.

**Case Aa**

In this case we have $\alpha \neq 0$, $\beta \neq 0$, the rank of the $6 \times 6$ Weyl projective curvature matrix is six and $\dot{R}^2 - R\ddot{R} = 0$. Equation $\dot{R}^2 - R\ddot{R} = 0 \Rightarrow R = be^{at}$, where $a,b \in R \setminus \{0\}$. The space-time (21) can, after a suitable rescaling of $x, y$ and $z$, take the form

$$ds^2 = -dt^2 + e^{2at}(dx^2 + dy^2 + dz^2). \tag{24}$$

It flows from [2,4] that the above space-times admits proper projective collineations. In this case Weyl projective curvature collineations are projective collineations.

**Case Ab**

In this case we have $\alpha \neq 0$, $\beta \neq 0$, the rank of the $6 \times 6$ Weyl projective curvature matrix is six and $\dot{R}^2 - R\ddot{R} \neq 0$. Subsitituting the above information into the Weyl projective equations (6) to (20) one finds $R = (at+b)^{-\frac{1}{3}}$, where $a,b \in R(a \neq 0)$ and the space-time becomes

$$ds^2 = -dt^2 + (at+b)^{-\frac{2}{3}}(dx^2 + dy^2 + dz^2). \tag{25}$$

Weyl projective curvature collineations in this case are

$$X^0 = a(at+b)^{-\frac{5}{3}} c_1, \quad X^1 = \frac{2}{3} x c_1 - y c_3 - z c_5 + c_7,$$

$$X^2 = \frac{2}{3} y c_1 + z c_2 + x c_3 + c_4, \quad X^3 = \frac{2}{3} z c_1 - y c_2 + x c_5 + c_6, \tag{26}$$

where $c_1, c_2, c_3, c_4, c_5, c_6, c_7 \in R$. One can write the above equation (26) after subtracting the Killing vector fields

$$X^0 = a(at+b)^{-\frac{5}{3}} c_1, \quad X^1 = \frac{2}{3} x c_1, \quad X^2 = \frac{2}{3} y c_1, \quad X^3 = \frac{2}{3} z c_1. \tag{27}$$

One can easily check that the above vector field is not proper curvature collineation because the existence of proper curvature collineation forces the rank of the $6\times 6$ Riemann matrix of the above space-time (25) to be three or less [2]. Here, the rank of the $6\times 6$ Riemann matrix is six. Hence the above space-time (25) does not admit proper curvature collineation. The situation is well clear when the above space-time adimts proper projective collineation [2,4]. Again the above space-time (25) does not admit proper projective collineation. The above vector field is not proper affine vector field because the existence of proper affine vector field forces the space-time to admit covariantly constant vector field and the rank of the $6\times 6$ Riemann matrix to be atmost three. Since the rank of the $6\times 6$ Riemann matrix is six. Hence the above vector field is not affine vector field. The situation is known when the above space-time admits proper affine vector field [6]. Lastly, one can easily show by writing the homothetic equations explicitly that the above vector field is not a proper homothetic vector field. Clearly, the above vector field (25) is a proper Weyl projective curvature collineation and it is finite dimensional.

**Case B**

In this case we have $\alpha \neq 0$, $\beta = 0$ and the rank of the $6\times 6$ Weyl projective curvature matrix is three. Equation $\beta = 0 \Rightarrow R = (at+b)^{\frac{1}{6}}$, where $a,b \in R(a \neq 0)$. The line element becomes

$$ds^2 = -dt^2 + (at+b)^{\frac{1}{3}}(dx^2 + dy^2 + dz^2). \qquad (28)$$

Subsitituting the above information into the Weyl projective curvature equations (6) to (20) we get one find that Weyl projective curvature collineations in this case are the Killing vector fields which are given in (22).

**Case C**

In this case we have $\alpha = 0$, $\beta \neq 0$, the rank of the $6\times 6$ Weyl projective curvature matrix is three and there exists a unique (up to a multiple) timelike vector field $t_a = t_{,a}$ satisfying $W^a{}_{bcd} t^d = 0$ but $t_{a;b} \neq 0$. Equation $\alpha = 0 \Rightarrow R = (et+d)^{\frac{2}{3}}$, where $e,d \in R(e \neq 0)$. The line element takes the form

$$ds^2 = -dt^2 + (et+d)^{\frac{4}{3}}(dx^2 + dy^2 + dz^2). \tag{29}$$

Subsitituting the above information into the Weyl projective curvature equations (6) to (20) and after some lenghty calculation one has

$$X^0 = \frac{3(et+d)}{e}c_5, \qquad X^1 = xc_5 - yc_6 + zc_7 + c_8,$$
$$X^2 = yc_5 + xc_6 - zc_9 + c_{11}, \quad X^3 = zc_5 - xc_7 + yc_9 + c_{10}, \tag{30}$$

where $c_5, c_6, c_7, c_8, c_9, c_{10}, c_{11} \in R$. The above equation (30) after subtracting the Killing vector fields becomes

$$X^0 = \frac{3(et+d)}{e}c_5, \quad X^1 = xc_5, \quad X^2 = yc_5, \quad X^3 = zc_5. \tag{31}$$

The Weyl projective curvature collineations in this case are homothetic vector fields.

**ACKNOWLEDGMENTS**

The author would like to thank the Abdus Salam International Centre for Theoretical Physics, Trieste, Italy for hospitality during his stay at the centre, where a part of this work was done. The author also thanks Prof. G S Hall (University of Aberdeen) for many helpful and interesting discussions.